\documentclass[preprint,12pt]{elsarticle}

\usepackage{amssymb}
\usepackage[intlimits,sumlimits]{amsmath}

\journal{Computer Physics Communications}

\begin{document}

\begin{frontmatter}



\title{Gaussian integration with rescaling of abscissas and weights}


\author{A. Odrzywolek}
\ead{odrzywolek@th.if.uj.edu.pl}
\address{M. Smoluchowski Institute of Physics, Jagiellonian University, Cracov, Poland}

\address{}

\begin{abstract}
An algorithm for integration of polynomial functions with variable weight is considered.
It provides extension of the Gaussian integration, with appropriate scaling
of the abscissas and weights. Method is a good alternative to usually adopted
interval splitting. 
\end{abstract}


\end{frontmatter}

\section{Introduction : the integral}
\label{}

In many areas of the physics and chemistry, the following integral
emerges:
\begin{equation}
\label{integral}
I(a) = \int_1^\infty \frac{g(x)}{1+e^{x/a}} \; dx
\end{equation}
where $g(x)$ is a smooth function and $0<a<\infty$ a real parameter. 
Here we assume that function $g(x)$ can be successfully
approximated by the polynomial. For a fixed value of $a$ the standard
method for numerical evaluation of such an integral is Gaussian integration
or similar closely related algorithm.

The difficulties caused by the integral of the form (\ref{integral})
can be show by comparison with other similar examples. If we change
lower integration limit from 1 to zero, one can easily remove
parameter $a$ from the algorithm by the substitution $x=a t$:
$$
a \int_0^\infty \frac{g(at)}{1+e^t} \; dt.
$$
Parameter $a$ now appear in the function $g$, and modified (because of 1 in the denominator)
Gauss-Laguerre algorithm can be used.

Similarly, if we remove 1 from the denominator, the substitution $z=(x-1)/a$ transform
integral into form:
$$
a e^{-1/a} \int_0^\infty g(1+az) e^{-z} \; dz
$$
easily integrable numerically with standard Gauss-Laguerre algorithm.

However, when both the lower integration limit is 1, and 1 is present in the denominator,
parameter $a$ cannot be eliminated from Eq.~(\ref{integral}): it always appears either
outside function $g$ in non-linear way, or enters the limit(s) of integration. 
This does not prevent us from calculating abscissas and weights of the
Gauss-like quadrature for any {\em fixed} value of $a$, say $a=1$, $a=3$ or $a=1/4$.
The general idea of the algorithm presented in this article, is to use Gauss-like
quadrature with abscissas and weights being functions of parameter $a$.

Standard method to handle integrals of the form (\ref{integral}) is to split
integration interval into at least two: $1 \leq x < \xi$, $\xi \leq x < \infty$.
If the value of $\xi$ is chosen properly as a function of $a$, then in the first
interval we have:
$$
\frac{1}{1+e^{x/a}} \sim \frac{1}{e^{\frac{1}{a}}+1}-\frac{e^{\frac{1}{a}} (x-1)}{a
   \left(e^{\frac{1}{a}}+1\right)^2}+O\left((x-1)^2\right),
$$
that is, weight is nearly linear (or polynomial if using higher order expansion).
In this interval we can use Gauss-Legendre integration. For $x>\xi$, 1 in the denominator
can be omitted, and Gauss-Laguerre quadrature apply. In practice, more interval
subdivisions are required \cite{2001CoPhC.136..294G, FXT_FD, 1998ApJS..117..627A}.

\section{Case with fixed parameter \label{sect:a1}}

We start analysis with simplest case of $a=1$ in Eq.~(\ref{integral}). 
The integral becomes:
\begin{equation}
\label{a_1}
I(1) =  \int_1^\infty \frac{g(x)}{1+e^{x}} \; dx
\end{equation}

For $g(x) = x^n$, where $n$ is non-negative integer, we have found\footnote{Although I do not
have a proof of this formula, it was verified using {\em Mathematica}
up to $n=23$.} that moments are equal to:
\begin{equation}
\label{a_1_xn}
M(n,1) \equiv  \int_1^\infty \frac{x^n}{1+e^{x}} \; dx = 
- \sum_{k=1}^{n+1} \left( \prod_{i=1}^{k-1} n+i-k+1 \right) 
\mathrm{Li}_k \left( -\frac{1}{e} \right)
\end{equation}
where $\mathrm{Li}_k$ is the polylogarithm. 
Our goal is to find weights, $W_i$, and abscissas, $X_i$, of the Gaussian quadrature
formula:
\begin{equation}
\label{orto1}
\int_1^\infty \frac{P_n}{1+e^{x}} \; dx = \sum_{i=1}^{N_{gauss}} W_i P_n(X_i),
\end{equation}
where $P_n$ is polynomial of the order up to $2 N_{gauss}-1$.

Convenient method \cite{GolubWelsh} to find weights and Gauss points uses orthogonal polynomials $Q_n$ related
to given weight:
$$
\int_1^\infty \frac{Q_n Q_m}{1+e^{x}} \; dx = \delta_{nm}.
$$
Unfortunately, as noted by \cite{Gautschi1970}, for infinite integration interval
calculations still require arbitrary precision calculations. Problem is increasingly
ill-conditioned numerically. We have found, that the best method is to solve system 
of equations for unknown polynomial coefficients in eq.~\eqref{orto1} progressively.
We use already found coefficients for polynomials of the order $n-1$, and
Laguerre polynomial $L_n$ coefficients as a guess starting points. System of the equations
is then solved numerically using arbitrary precision arithmetic. Results were verified
using method of \cite{Fukuda2005143}.

Once orthogonal polynomials are found, abscissas are zeros of the $Q_n$, and weights
are equal to \cite{Gautschi1970}:
$$
W_i = \left ( \sum_{k=1}^{n-1} Q_k(X_i)^2 \right)^{-1}.
$$
Another equivalent formula for weights is \cite{GolubWelsh}:
$$
W_i = -\frac{k_n}{k_{n-1}} \left( Q_n(X_i) Q_{n-1}^{'}(X_i) \right)^{-1}
$$

Knowledge of abscissas and weights for integral with $a=1$ is crucial, 
because of the approximate scaling found and used in the next section
to derive more general results for $a \neq 1$.

\section{Scaling of abscissas and weights}

For integral (\ref{integral}) with function $g(x)$
being equal to $x^n$ we can provide formula similar to (\ref{a_1_xn}), 
generalized to case $a \neq 1$:
\begin{equation}
\label{a_xn}
M(n,a) = \int_1^\infty \frac{x^n}{1+e^{x/a}} \; dx = 
- \sum_{k=1}^{n+1} a^k \, \left( \prod_{i=1}^{k-1} n+i-k+1 \right) 
\mathrm{Li}_k \left( -\frac{1}{e^{1/a}} \right)
\end{equation}

Knowledge of the moments facilitates calculations of abscissas
and weights, but in practice polylogarithms present in eq.~\eqref{a_xn}
often is not directly available. One must calculate moments by means 
of direct numerical integration. In principle, using \eqref{a_xn},
one can find orthogonal polynomials (and Gaussian quadrature abscissas
and weights as well) analytically. However, formulae become ridiculously
complicated already for $n>2$, and we restrict discussion to case
of $n=1$. 

Unfortunately, we were unable to derive general formulae for orthogonal
polynomials $Q_n$ or find coefficients of the three-term recurrence formula.

\subsection{Single Gaussian point quadrature}

\begin{figure}
\includegraphics[width=0.95\textwidth]{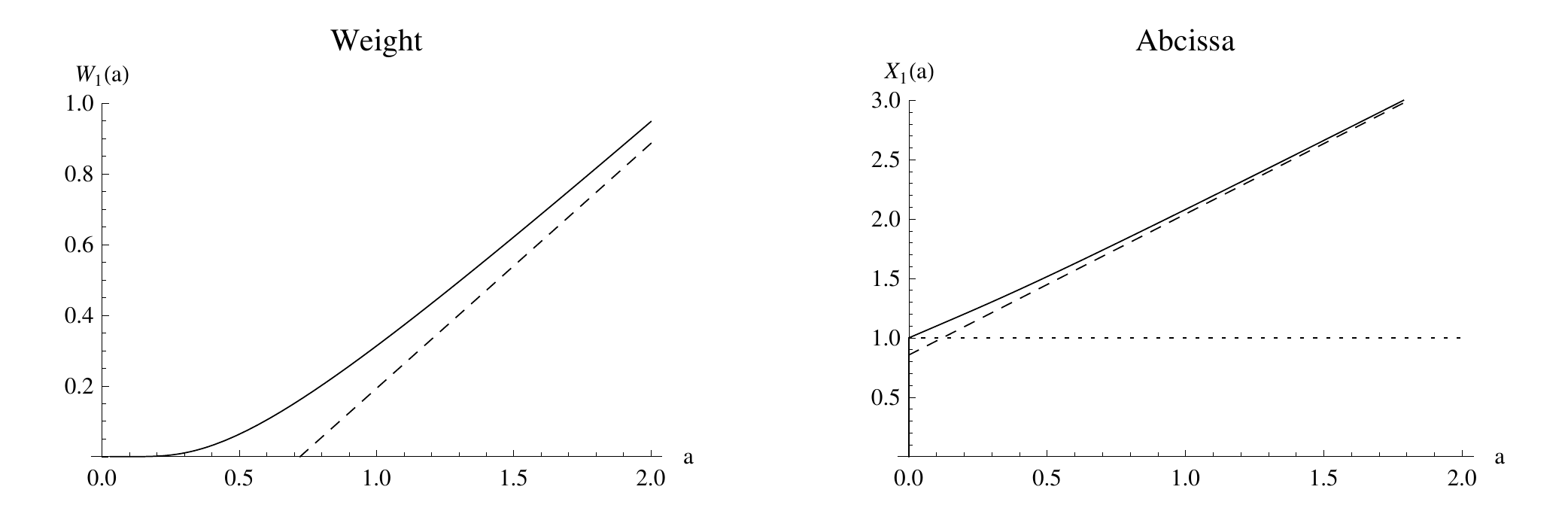}
\caption{\label{W1X1} Weight (left) and abscissa (right) and  for single point Gaussian quadrature
as a functions of $a$, and its asymptotes (dashed).
}
\end{figure}

While the case of single Gaussian point is not
interesting from practical point of view, it gives
some insights into planned procedure due to very simple formulae.
Quadrature is:
\begin{equation}
\int_1^\infty \frac{g(x)}{1+e^{x/a}} \; dx  \simeq w_1  g(x_1),
\end{equation}
where:
\begin{equation}
w_1 = a \ln{(1+e^{-1/a})} \equiv a\; \mathrm{Li}_1 (-e^{-1/a}), \quad 
x_1 = 1+a\; \frac{\mathrm{Li}_2 (-e^{-1/a})}{\mathrm{Li}_1 (-e^{-1/a})}
\end{equation}
Integration algorithm based on abscissa and weight provided above is exact only for constant functions. 
Weight and abscissa are shown
in Fig.~\ref{W1X1}. Functions
$\mathrm{Li}_k (-e^{-1/a})$ present in the formulas above are somewhat pathological.
All derivatives vanish at $a=0$:
$$
\lim_{a \to 0^+} \frac{d^n}{da^n} \mathrm{Li}_k (-e^{-1/a}) = 0.
$$
Therefore, the function $\psi(a)$ defined as follows:
\begin{equation}
\label{psi}
\psi(a)=
\begin{cases}
0 & a \leq 0 \\
\mathrm{Li}_k (-e^{-1/a}) & a>0
\end{cases}
\end{equation}
is an example of the infinitely differentiable function which is non-analytic
at $a=0$. Therefore, it cannot be approximated by the polynomials
near $a=0$. For $a \to \infty$ we have:
\begin{equation}
\lim_{a \to \infty} \mathrm{Li}_k (-e^{-1/a}) =
\begin{cases}
- \ln{2} & \quad k=1\\
\left( 4^{(1-k)/2} - 1 \right) \, \zeta(k) & \quad k>1
\end{cases}
\end{equation}
where $\zeta$ is Riemann-zeta function. Both functions $W_1(a)$ and $X_1(a)$ approach
their asymptotes very slowly, cf. Fig.~\ref{W1X1}.

\subsection{General Gaussian quadrature}

 For $a\neq1$, due to extremely complicated formulae, discussion will be limited to
 numerical results. It is observed, that functions $W_i(a)$ and $X_i(a)$ behave
 like functions $W_1(a)$ and $X_1(a)$ discussed in previous subsection.
 Rough approximation for abscissas and weights (arbitrary order) can be combined from results
 for $a=1$, and single point quadrature as follows:
\begin{equation}
\label{simple_scalling}
 X_i(a) \simeq 1 + a \, \lambda_i, \qquad W_i(a) \simeq \delta_i\, (-1 + a \ln{(1+e^{1/a})})
\end{equation}
 where $\lambda_i = 1+ X_i(1)$ and $\delta_i = W_i(1)/(-1+\ln{(1+e)})$. Values
 of $W_i(1)$ and $X_i(1)$ are simply weights and abscissas for $a=1$ quadrature, 
 cf. Sect.~\ref{sect:a1}.

\begin{figure}
\includegraphics[width=\textwidth]{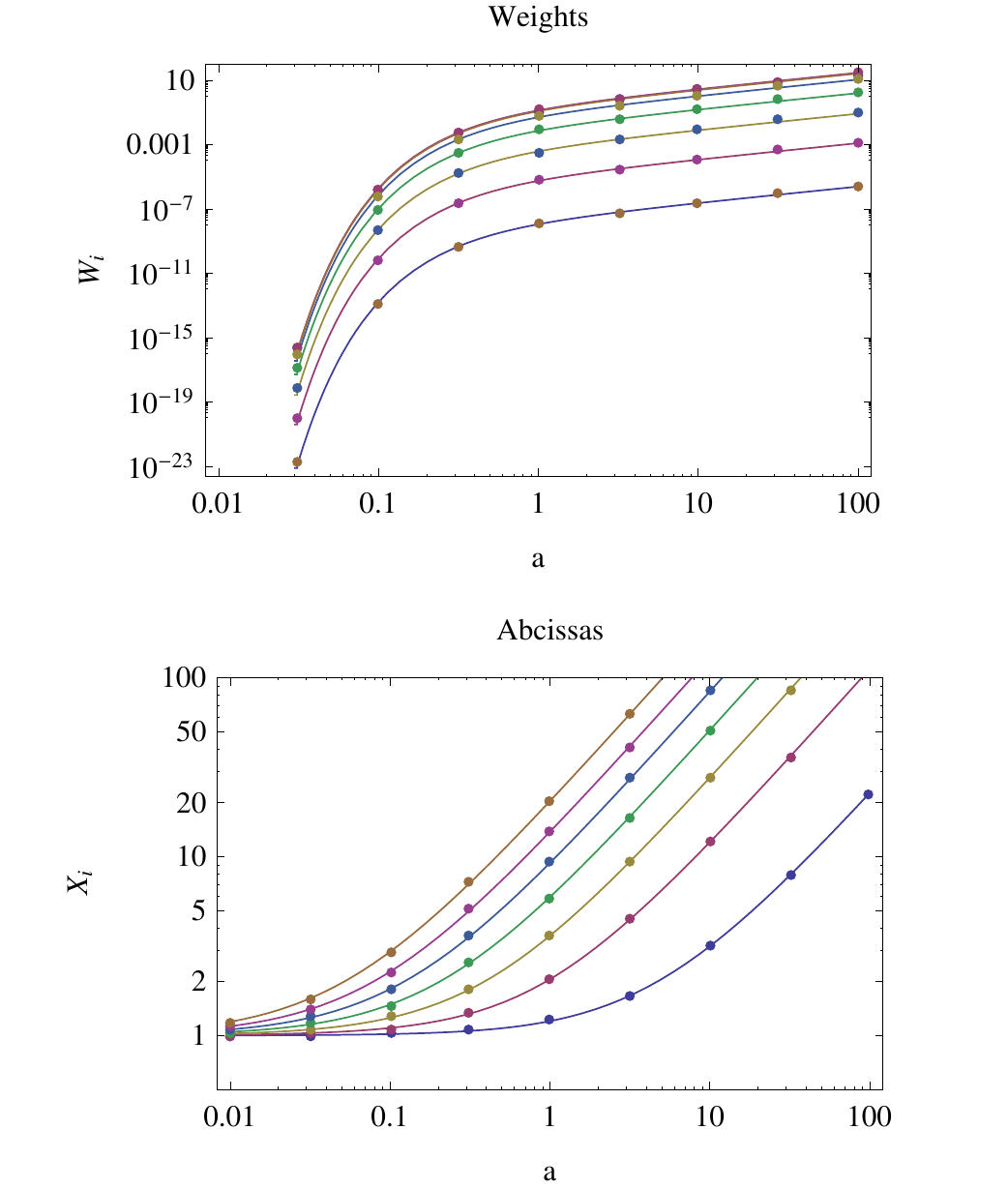}
\caption{\label{WaXa} Numerically obtained Gaussian weights and abscissas
for integral of the form \eqref{integral} (points) compared to formulae 
\eqref{simple_scalling} (lines). 7-point quadrature is shown.
}
\end{figure}

 Example results are presented in Fig.~\ref{WaXa} using lines.
 Gauss points behave as expected for $a \to 0$, that is, they concentrate near $x=1$.
 For $a \to \infty$ abscissas spread from $x=1$ to infinity.
 Weight also behave correctly, scaling linearly with $a \to \infty$ and approaching
 zero as $a \to 0$. For $a \to 0$, formulae \eqref{simple_scalling} are progressively
 better approximations. Note however, that \eqref{simple_scalling} do not provide
 correct asymptotic behavior for $a \to \infty$. This would require knowledge
 of analytical formulae for $W_i(a)$ and $X_i(a)$, or its asymptotic expansion at
 least.

Parameterization \eqref{simple_scalling} provides excellent starting guess 
points\footnote{Without guess points it is still possible to find $X_i(a), W_i(a)$
starting with known values $X_i(1), W_i(1)$, slowly changing $a$ and using previously
obtained values in the next step as an another guess. However, this procedure is inherently
{\em sequential}, while using parameterized guess one can do all calculations
{\em in parallel}. Mathematica script of \cite{Fukuda2005143} also
do not need any guess points, but require 500 seconds to find 10-point quadrature 
for \eqref{a_1}. Our method do the same job in 2 to 3 seconds on the same machine.
}
for numerical calculations of $W_i(a)$ and $X_i(a)$.

Typical numerical result is shown in Fig.~\ref{WaXa} using bullets. 
Apparent accuracy of the scaling
\eqref{simple_scalling} is misleading, 
because (i) logarithmic scale hide errors (ii) very high accuracy is required
for Gaussian quadrature to work. Therefore quadrature obtained from
\eqref{simple_scalling} is not expected to be very accurate, except maybe
for $a \ll 1$. Nevertheless, overall picture is appealing, and use of scaling, 
possibly not as simple as \eqref{simple_scalling},
seems to be move in the promising direction. At present, however, we are forced
to use interpolation of numerically computed values. For values of $a$ outside
interpolation domain we can use asymptotes for $a \gg 1$, and \eqref{psi} for $a \ll 1$.

\section{Performance of the algorithm}

Approximate value of the integral \eqref{integral} is computed from formula:
\begin{equation}
\label{algorithm}
I(a) \simeq  \sum_{i=1}^{N_{gauss}} W_i(a) P_n \left( X_i(a) \right)
\end{equation}
where, in contrast to \eqref{orto1}, $W_i$ and $X_i$ are functions of $a$, 
cf. Fig.~\ref{WaXa}.

 There are two factors determining accuracy of the method: number of Gaussian
 points, $N_{gauss}$, and accuracy of the functions $W_i(a), X_i(a)$, 
 $i=1 \ldots N_{gauss}$. The latter depends
 on number of sampling points, $N_{sampl}$, and algorithm used to interpolate between them.
 Ideally, we would like to have many Gauss points, and very accurate 
 (or exact) scaling of them. In practice we should know what is better:
 large $N_{gauss}$ with small $N_{sampl}$, or {\em vice versa}. We address these questions
 in next subsections.

\subsection{Accuracy as a function of Gaussian points  $N_{gauss}$ }

In this subsection we keep number of sampling points used to interpolate
functions $W_i(a)$ and $X_i(a)$ fixed. Number of these functions will be varied.
Naively, from definition of the Riemann integral, we expect that more sampling
points could result in increased accuracy regardless of the algorithm used, as long
as integral is convergent. On the other hand Gaussian quadrature of polynomials 
is exact up to certain order, but only
for precisely determined abscissas and weights. Algorithm is then expected to
fail (in the sense of accuracy) for polynomials except at the grid points. For non-polynomial functions
Gaussian integration is only approximate functional, and accuracy of
$W_i, X_i$ is possibly less important.

Now we attempt to test these expectations using two examples of interest:
\begin{subequations}
\label{tests}
\begin{equation}
\label{test1}
g_1(x) = x+x^4-x^5, \qquad \text{polynomial exactly integrable}
\end{equation}
\begin{equation}
\label{test2}
g_2(x) = x^4 \sqrt{1+x}, \qquad \text{function arising from relativistic momentum}
\end{equation}
\end{subequations}

Number of sampling points has been fixed to 50 per decade per function. In the range
of $0.01 \leq a \leq 100$ total of 201 points spaced logarithmically were used.
Third and eight order interpolation were used for abscissas and weights, respectively.

\begin{figure}
\includegraphics[width=\textwidth]{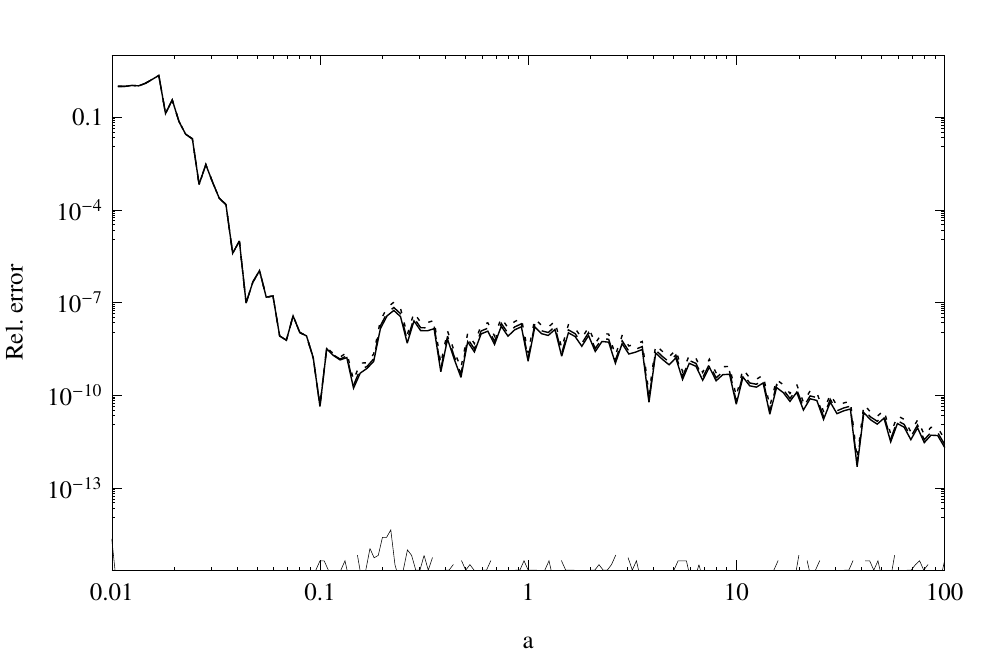}
\caption{\label{RelAccuracy_Ngauss_poly} Relative errors versus $a$ 
for $N_{gauss}=3$ (dotted), $5$ (dashed) and $11$ (solid) for polynomial \eqref{test1}.}
\end{figure}

Three cases are compared: $N_{gauss}=3, 7$ and $11$. 
Relative accuracy obtained for polynomial 
\eqref{test1} is presented in Fig.~\ref{RelAccuracy_Ngauss_poly}.
Differences are barely visible, and error is entirely due to
variations of the abscissas and weights. Normally, for polynomial
of the order $\leq5$ we expect to achieve relative error of the order
of machine epsilon (Fig.~\ref{RelAccuracy_Ngauss_poly}, thin line near bottom;
axis is placed at $\epsilon=2.2 \times 10^{-16}$).
Therefore, this figure shows errors caused by the interpolation of
the abscissas and weights. Error estimate reach huge values
for small values of $a$. However, in typical application integrals
of the form \eqref{integral} are calculated over wide range of
parameters and summed up. Therefore overall absolute error is mainly
due to regions with large $a$ where the numerical value of 
the integral is largest. Interpolation error visible in Fig.~\ref{RelAccuracy_Ngauss_poly} 
might be significantly reduced if required, 
cf. Sect.~\ref{sect:Nsampl} and Figs.~\ref{RelAccuracyNpoints}, \ref{RelAccuracyOrder}.

\begin{figure}
\includegraphics[width=\textwidth]{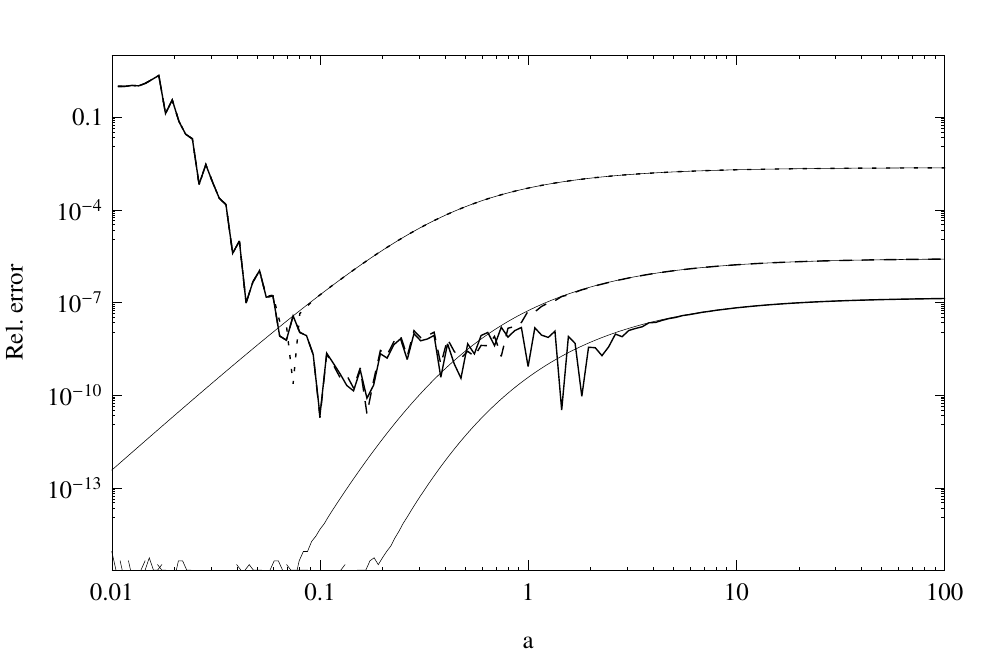}
\caption{\label{RelAccuracy_Ngauss_nonpoly} Like in Fig.~\ref{RelAccuracy_Ngauss_poly},
but for non-polynomial function \eqref{test2}.}
\end{figure}

Test case $g_2$ \eqref{test2} provides more realistic task for the algorithm.
This time number of Gaussian points do matter, especially for $a>1$. Thin lines
show relative accuracy achieved exactly at grid 
points\footnote{To get this visual effect, we use original grid to
draw thin lines, and irrational step to draw thick lines.}. For $a>1$,
algorithm surprisingly provides accuracy nearly identical to normal
Gaussian integration. It is however clear from Fig.~\ref{RelAccuracy_Ngauss_nonpoly},
that accuracy better than $\sim 10^{-10}$ cannot be achieved, regardless of the 
value of $N_{gauss}$. We need more dense grid of points used to interpolate functions
$W_i(a), X_i(a)$ in \eqref{algorithm}, 
or better interpolation algorithm, cf. Sect.~\ref{sect:Nsampl}. Again, for $a \ll 1$
relative accuracy is poor, but integral values are very small 
in this regime due to exponential
factor in the integrand of \eqref{integral}. For small $a$ one can also consider
use of \eqref{simple_scalling} instead of numerical values, 
see Fig.~\ref{RelAccuracyNpoints}, red line.

\subsection{Accuracy as a function of sampling points $N_{sampl}$ \label{sect:Nsampl}}

In this subsection we keep number of Gaussian points fixed, \mbox{$N_{gauss}=7$}.
Number of sampling points will be increased, so interpolated functions
$W_i(a), X_i(a)$ will approach their exact values. For polynomials up to the
order $2 N_{gauss}-1 = 13$, integration algorithm 
is expected to achieve machine precision -- in limiting case at least.
For other functions it should be no worse than ordinary Gauss-Laguerre-like
integration with fixed weight.

\begin{figure}
\includegraphics[width=\textwidth]{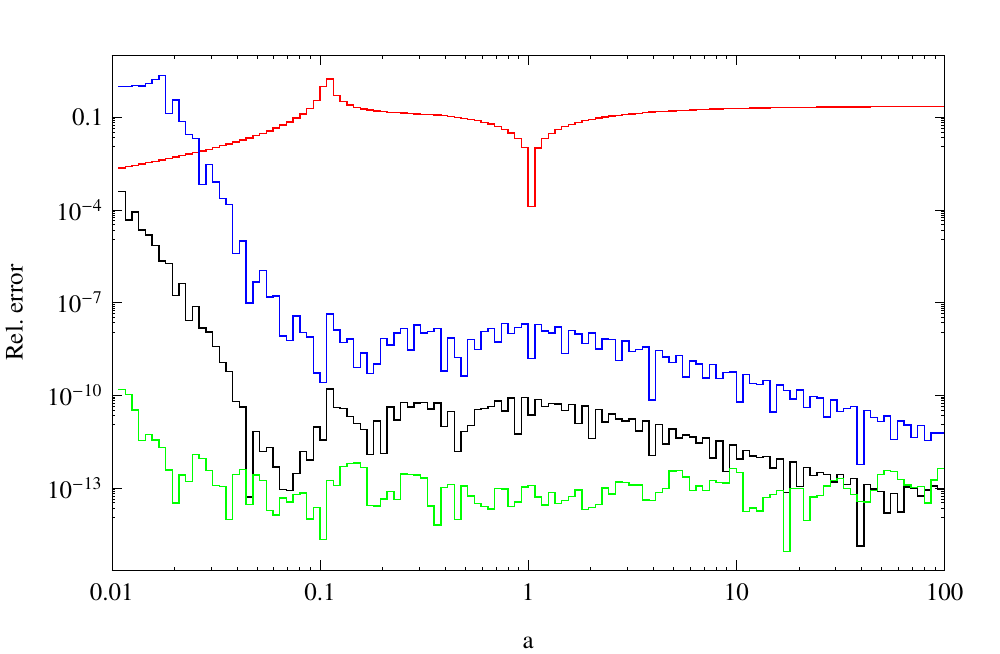}
\caption{\label{RelAccuracyNpoints} Relative errors versus $a$ caused in the final integral 
using increased number of interpolation points used to restore abscissas and weights 
for integrand \eqref{test1}. 
Red - analytical formula, blue - 50/decade, black -200/decade,
green - 1000/decade.
}
\end{figure}

Typical result is presented in Fig.~\ref{RelAccuracyNpoints}. Let us remind, that
in the case of exact abscissas and weights for polynomial \eqref{test1} we expect
to achieve machine precision. In Fig.~\ref{RelAccuracyNpoints} 
$\epsilon=2.2 \times 10^{-16}$, where lower axis has been placed. 
Third order interpolation has been used
do compute abscissas and eight order to compute weights. Analytical formula \eqref{simple_scalling} is shown
in red for reference. It is not surprise, that using progressively more 
interpolation points we get better results (Fig.~\ref{RelAccuracyNpoints}). Using
1000 points per decade (Fig.~\ref{RelAccuracyNpoints}, green), relative accuracy
is at level of $10^{-13}$.

\subsubsection{Influence of the interpolation order and method}

\begin{figure}
\includegraphics[width=\textwidth]{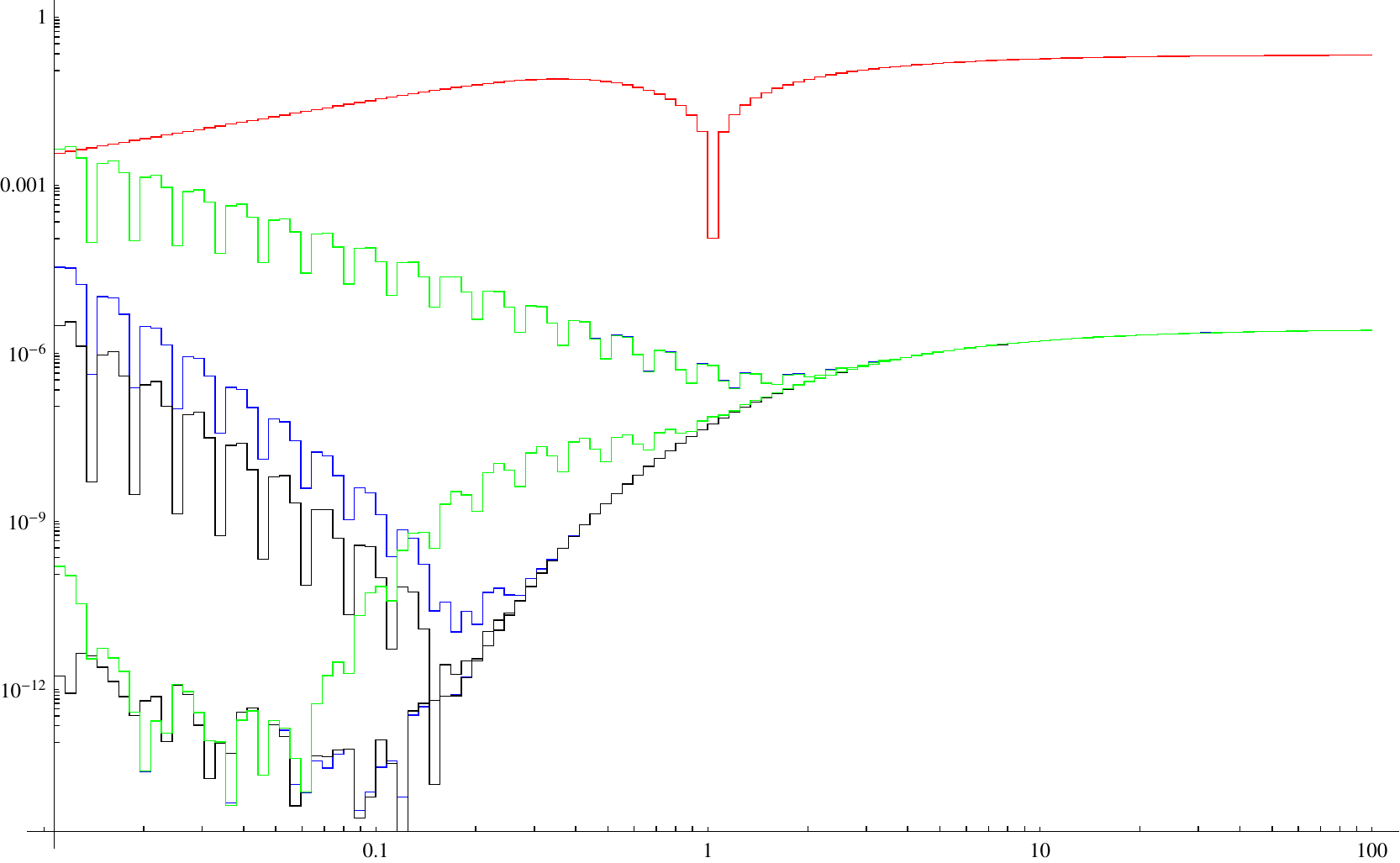}
\caption{\label{RelAccuracyOrder} Relative errors versus $a$ caused in the final integral 
using various interpolation
algorithms used to restore abscissas and weights (using 1000 points per decade) 
for integrand \eqref{test2}. 
Red - analytical formula, blue - Hermite, black - splines, 
green - mixed orders (see text).
}
\end{figure}

Interpolation goal is to reproduce abscissas and weights from discrete set of points
as accurately as possible. Besides
number of these points, which is obvious factor, secondary important issues are:
interpolation method, interpolation order and location of the points.
Influence of various factors on relative integration error in the case of function 
\eqref{test2}
is presented in Fig.~\ref{RelAccuracyOrder}. 4000 points spaced 
logarithmically has been used in the range from 0.01 to 100, i.e. thousand points per
decade. Two standard interpolation algorithms: Hermite and spline were used
with order 1 (linear), 3 and 8. Noteworthy, relative error gain obtained from increase of 
the interpolation order from 3 to 8 is as high as six orders of magnitude. 
Further test has shown, that gain is almost entirely due to accuracy of weights. Abscissas
are successfully approximated even using linear interpolation, cf. green lines
in Fig.~\ref{RelAccuracyOrder}. Interpolation order and method
is important mainly if $a<1$. For $a \gg 1$ no differences are visible.

Possibly, using non-linearly scattered points and renormalized function values
we would be able to increase accuracy further without increasing number of points.
This might be important issue in real-world implementation of the algorithm. Too large
amount of data can result in cache misses of modern CPUs, slowing down calculations.

\section{Concluding remarks}

Novel algorithm for evaluation of the improper integrals with a real parameter \eqref{integral}
in the weight function has been presented. Method is based on Gaussian integration,
where abscissas and weights are functions of parameter $a$ rather than just real numbers.
Both analytical and numerical approximations of these functions are considered.
Algorithm is shown to be robust and useful.

Presented method might be considered to be an example of rendering of the multivariate
function into number of functions of one variable. The latter can be approximated 
very accurately, and do not consume computer memory, in contrast to tabulations
of multivariate functions. Therefore the fact, that functions $W_i$ are not
already known in terms of e.g. elementary functions is not serious limitation
for modern hardware.

Possible direction of further research are: (i) search for exact scaling formulae
or their approximations (ii) scaling in limiting cases $a \to 0$ and $a \to \infty$
(iii) dedicated interpolation formulae (iv) use of scaling with non-Gaussian (e.g. trapezoidal)
integration rules. Properties of the related orthogonal (non-classic) polynomials 
are also of interest, particularly three-term recurrence formula and 
behavior of the leading terms.

\section*{Acknowledgements}

The research was carried out with the supercomputer Deszno purchased thanks
to the financial support of the European Regional Development Fund in the framework
of the Polish Innovation Economy Operational Program (contract no. POIG.
02.01.00-12-023/08). I would like to thank students who struggled to solve
many related tasks during \textit{Advanced Symbolic Algebra} course taught at
the Jagiellonian University in Cracov 2009/2010.

\bibliographystyle{cpc}
\bibliography{GaussScallingIntegration}

\begin{thebibliography}{1}

\bibitem{2001CoPhC.136..294G}
{Gong}, Z., {Zejda}, L., {D{\"a}ppen}, W., and {Aparicio}, J.~M.,
\newblock Computer Physics Communications {\bf 136} (2001) 294.

\bibitem{FXT_FD}
{Timmes}, F.,
\newblock \mbox{Cococubed.com},
\newblock http://cococubed.asu.edu/code\_pages/fermi\_dirac.shtml, 2008.

\bibitem{1998ApJS..117..627A}
{Aparicio}, J.~M.,
\newblock Astrophysical Journal Supplement {\bf 117} (1998) 627.

\bibitem{GolubWelsh}
{Golub}, G.~H. and H., W.~J.,
\newblock Mathematics of Computation {\bf 23} (1969) 221.

\bibitem{Gautschi1970}
{Gautschi}, W.,
\newblock Mathematics of Computation {\bf 24} (1970) 245.

\bibitem{Fukuda2005143}
Fukuda, H., Katuya, M., Alt, E., and Matveenko, A.,
\newblock Computer Physics Communications {\bf 167} (2005) 143 .

\end{thebibliography}

\end{document}